\documentclass{aa}  
\usepackage{multirow}
\usepackage{amsmath}	
\usepackage{amssymb}
\usepackage{graphicx}
\graphicspath{{figs/}}
\usepackage{comment}
\usepackage{txfonts}
\allowdisplaybreaks
\usepackage{hyperref}
\hypersetup{
	pdftitle={On a discontinuity at the base of the transition layer located between the  Keplerian accretion disk and the compact object},
	pdfsubject={Astrophysics...},
	pdfauthor={L. Titarchuk, I. Kalashnikov}
}

\usepackage{natbib}
\bibpunct{(}{)}{;}{a}{}{,} 

\usepackage{xcolor}
\newcommand{\ik}[1]{\textcolor{black}{#1}}
\newcommand{\col}[1]{\textcolor{black}{#1}}

\usepackage[pstricks]{bclogo}

\begin{document}

   \title{On a discontinuity at the base of the transition layer located between the  Keplerian accretion disk and the compact object}

   \author{L. Titarchuk
          \inst{1}
          \and
          I. Kalashnikov\inst{2}
          }

   \institute{Dipartimento di Fisica, Universit\`a di Ferrara, via Saragat 1, 44122 Ferrara, Italy\\
              \email {titarchuk@fe.infn.it }
         \and
             Keldysh Institute of Applied Mathematics,  Russian Academy of Sciences. 4 Miusskaya sq., Moscow, 125047, Russia\\
             \email{kalasxel@gmail.com}
             }
         
   \titlerunning{On a discontinuity at the base of the transition layer\ldots}

   \date{Received; accepted.}

  \abstract
{We study the geometry of the transition layer (TL) between the classical Keplerian accretion disk (the TL outer  boundary) and the compact object at the TL inner boundary. }
  { 
  Our goal is to demonstrate using the hydrodynamical formalism that the TL is created along with a shock  due to a discontinuity and  to an adjustment of the Keplerian disk  motion  to a central object. }
   {We apply hydrodynamical equations to  describe a plasma motion  near a central object in the TL.  }
   { We point out that before matter accretes to a central object the TL cloud is formed  between an adjustment radius and the TL inner boundary which is probably a {site} where the emergent Compton spectrum  comes from. \col{Using a generalization of  the Randkine-Hugoniot relation and a solution of the  azimutal force balance equation 
   we have reproduced the geometric characteristics of TL. }  }
   {}
   
  { 
   }
   { }
   {  }
   {}


   \maketitle
%

\section{Introduction}\label{sec:Intro}
Accretion onto compact objects  as white dwarf (WD), neutron star (NS), and black hole (BH) binaries are very similar. The accreting matter with relatively large angular momentum  can form  a disk  or go to outflow  (see \cite{Titarchuk07} and \cite{Shaposhnikov09}, hereafter TSA07, ST09, respectively).   On the other hand, the plasma with low angular  momentum  proceeds towards  the compact object almost  in a free-fall manner  until  the centrifugal force 
becomes strong  to halt the flow (see, e.g., \citep{Chakrabarti95}, hereafter CT95) Thus, we can suggest  that  three definitive regions  a (Keplerian) disk, the  shock, and  a transition layer (TL)  between the last stable orbit near {WD}, NS or BH and the  shock, which are probably a place where the emergent X-ray spectrum is formed.  In this paper, we study  the characteristics of this adjustment of the  Keplerian disk  flow  to   an {WD,}  an NS or an BH. 

The disk starts to deviate from a Keplerian motion at a certain radius to adapt itself to the boundary conditions at  WD or NS surface (or at the last stable orbit, at \ik{$r_1 = 3r_{\rm S}$, where $r_{\rm S}$} is the Schwarzchild radius, in the case of a BH).  
\cite{Titarchuk98} (hereafter TLM98) explained the millisecond variability detected by {Rossi} X-Ray Timing Explorer ({RXTE}) in the X-ray emission from a number of low-mass X-ray binary systems (XRBs). Later,  \cite{Seifina10},  \cite{Farinelli11},  and so on analyzed X-ray data from Sco X-1, 4U 1728$-$34, 4U 1608$-$522, 4U 1636$-$536, 4U 0614$-$091, 4U 1735$-$44, 4U 1820$-$30, GX 5$-$1) in terms of dynamics of the centrifugal barrier,  in a hot boundary region  (the TL) surrounding a neutron star (NS). They  demonstrated that this region may experience relaxation oscillations and that the displacements of a gas element both in radial and vertical directions occur at the same main frequency, of an order of the local Keplerian frequency.

Observations of black hole X-ray binaries and active galactic nuclei indicate that the accretion flows around black holes are composed of hot and cold gas, which have been theoretically described in terms of a hot  corona next to an optically thick relatively cold disk. \cite{Liu22} (hereafter LQ22)  reviews the accretion flows around black holes, with an emphasis on the physics that determines the configuration of hot and cold accreting gas, and how the configuration varies with the accretion rate and thereby produces various luminosity and spectra. They provide references to the famous solutions of the standard disk model such as, for example, \cite{Shakura73} (hereafter SS73). The most successful applications of this model are the steady state and time-dependent nature of thin disks in dwarf novae and soft X-ray transients. Application of the four accretion models to black holes presented by  LQ22  is constrained by both the theoretical assumption of specific models and observational luminosity and spectrum. 

The  soft photons  from the innermost part of the disk  are scattered off the hot electrons thus forming the Comptonized X-ray spectrum [see \cite{Sunyaev80}, hereafter ST80]. The electron temperature is regulated by the supply of soft photons from a disk, which depends on the ratio of the energy release (accretion rate) in a disk and the energy release in the TL. For example, the electron temperature is higher for lower accretion rates (see, e.g., CT95), while for a high accretion rate (of an order of the Eddington one) the TL
is cooled down very efficiently due to the soft photon flux and the Comptonization. {This is a possible mechanism} for the low-frequency QPOs (quasi-periodic oscillations), discovered by RXTE in a number of low-mass X-ray binary systems (LMXBs; see \cite{Strohmayer96,vanderKlis96}, hereafter S96 and VK96a, respectively; \cite{Zhang96}, TLM98, \cite{Titarchuk07} \ik{hereafter} TSA07). These observations reveal a wealth of  high and low-frequency X-ray variabilities that are believed to be due to the processes occurring in the very vicinity of an accreting NS and BH.

\col{
\cite{Lancova19} discovered a new class of black hole accretion disk solutions through 3D radiative magnetohydrodynamic simulations applying  the general relativity formalism. These solutions combine features of thin, slim, and thick disk models. They provide a realistic description of black hole disks and have a thermally stable, optically thick, Keplerian region supported by magnetic fields. \cite{Mishra20} have conducted a detailed analysis of two-dimensional viscous, radiation hydrodynamic numerical simulations of  the Shakura-Sunyaev thin disks around a stellar mass black hole. They found multiple robust, coherent oscillations occurring in the disk, including a trapped fundamental g-mode and standing-wave p-modes.
The study suggests that these findings could be of astrophysical importance in observed twin peak, high-frequency QPOs. Large scale 3D magnetohydrodynamic simulations of accretion onto magnetized stars with tilted magnetic and rotational axes were done by \cite{Romanova21}.It was shown that the inner parts of the disc become warped, tilted, and precess due to magnetic interaction between the magnetosphere and the disc. According to results of the numerical simulations by \cite{Lugovskii12}, large-scale instability in accretion disks, reveals changes in a disk flow structure because of  t the formation of large vortexes. Over time, this leads to the development of asymmetric spiral structures and results in angular-momentum redistribution within the disk. }

\col{
\cite{Sukova17} have conducted simulations of accretion flows with  a low angular momentum, filling the gap between spherically symmetric Bondi accretion and disc-like accretion flows. They identify ranges of parameters for which the shock after formation moves towards or outwards the central black hole or the long-lasting oscillating shock is observed. The results are scalable with a central black hole mass  and can be compared to QPOs of selected microquasars and supermassive black holes in the centres of weakly active galaxies. In order to explain {NICER} telescope evidences for non-dipolar magnetic field structures in rotation-powered millisecond pulsars, \cite{Das22} have conducted a suite of general relativistic magnetohydrodynamic simulations of accreting neutron stars for dipole, quadrupole, and quadrudipolar stellar field geometries. The study found that the location and a size of the accretion columns resulting in hotspots changes significantly depending on initial stellar field strength and geometry, providing a viable mechanism to explain the radio power in observed neutron star jets.
}

Despite the increasing role of numerical simulations of accretion, the consideration of relatively simple analytical models allows us to identify the response of the system to changes in a set of parameters, which in the case of numerical simulations becomes a very cumbersome task. In the present paper we, along with other authors (e.g.\@ \cite{Abramowicz13,Ajay22})), propose a model of accretion on the central object. 
\textcolor{black}{The point is that existing models of shock waves (see e.g.\@ \cite{Chakrabarti89}) cannot correctly describe the abrupt transition between the accretion disk and the TL. Our goal is to construct a satisfactory model of the shock wave lying at the base of TL, which can correctly describe observational  data for  temperature and density (\cite{Seifina10}).}
In Sect.~\ref{motivation_model} we describe our approach to the  TL model. 
In Sect.~\ref{vert_depend}. we formulate equations  to determine  the vertical dependencies of density and pressure in the TL. In Secs.~\ref{R-H}-\ref{rot} we introduce  the generalized  Rankine-Hugoniot relation and take into account the rotation effect in the TL, respectively.
We discuss our setup and final results   in Sect.~\ref{setup}.
Finally, we summarize our conclusions in Sect.~\ref{conclusion}.

\section{Motivation and the model}
\label{motivation_model}
\begin{figure}
	\centering\includegraphics[width=1\linewidth]{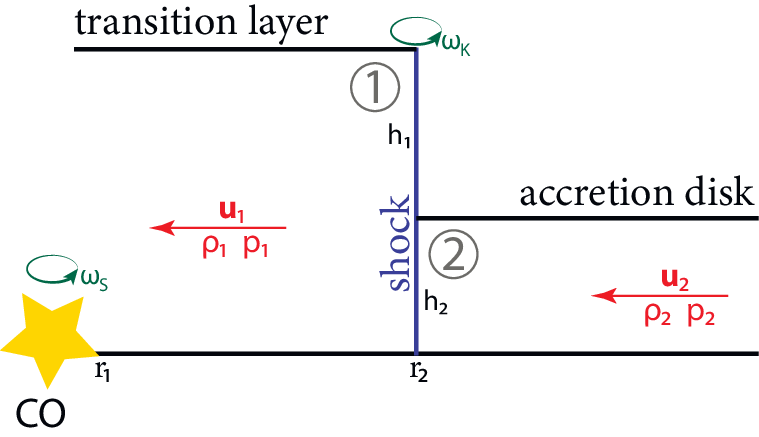}
	\caption{The schematic representation of the considered model.}
	\label{fig:scheme}
\end{figure}

Consideration of the two-dimensional structure of accretion disks is associated with obvious mathematical difficulties. Typically, a two-dimensional structure is only accounted for by an non-homogeneous distribution of matter within a layer of constant thickness $2h$. Consideration of a disk of constant thickness is made possible by ignoring the vertical velocity $v_z$. The more general case $h=h(r)$ requires solving equation $dh/dr = (v_z/u)|_{z=h(r)}$, where $u$ is the radial velocity, to describe the behavior of the height as the radius changes. 

According to the observations and their interpretations (TLM98, ST09)  the disk height may achieve very significant values near the TL, while at larger distances it is pretty small. The thermodynamic characteristics of the plasma change considerably: in the TL, its temperature can reach tens of keV and its density drops drastically (see e.g. ST09). This change is widely believed  to be presumably due to the presence of a shock wave (see e.g. CT95, TLM98 and ST09).
 Such a region near the central object (CO) is  associated with  the TL.  

In an attempt to get closer to a two-dimensional consideration, we have chosen the following model describing the shock (Fig.~\ref{fig:scheme}). The flow velocity is supposed to have only radial and azimuthal components in the both sides of the shock, which are independent of the height. It is assumed that the height of the flow changes sharply by discontinuity from $h_2$ (the disk thickness) to $h_1$ (the TL thickness).

We require the angular velocity $\omega$ of the TL to match to the CO at its surface and coincide with disk velocity at the shock position $r_2$. This angular velocity of the disk is supposed to equal to Keplerian one.
Only $r\phi$ component of the viscous stress tensor was considered because it is responsible for taking off the angular momentum. We considered the stress tensor form $\eta r \omega \omega'$, where $\eta$ is the turbulent viscosity.

The thermodynamic quantities of the flow have vertical dependencies which are determined by the CO gravity and an equation of state. In order to identify them, we did not assume the smallness of the vertical coordinate compared to the radial one.

\section{The vertical dependences}
\label{vert_depend}
Let us start with the hydrostatic balance equation in the vertical direction: 
\begin{align}
	\frac{\partial p}{\partial z} = - \rho \frac{\partial\Phi}{\partial z},
	\label{vertBal}
\end{align}
where $p$,  $\rho$ are pressure and  density and $\Phi$ is the gravitational potential creating by the central object. We supposed that the vertical distribution of matter is given by polytropic relation $ p=K \rho^{1+1/n}$, where $K$ is a constant related to the entropy and $n$ is the polytropic index. Then the formal solution of (\ref{vertBal}) may be written as:
\begin{align}
	K (n+1) \rho^{1/n} = \widetilde{f}(r) - \Phi(r,z),
\end{align}
where $\widetilde{f}$ is an arbitrary function. We require the density and pressure to be zero at the inner boundary $z=\pm h$ and equal to some functions $\rho_e$, $p_e$ correspondingly at the equator $z=0$. Then we have:
\begin{align}
	& \rho = \rho_e(r) \left( \frac{\Phi(r,h)-\Phi(r,z)}{\Phi(r,h)-\Phi(r,0)} \right)^n, \label{rhoDep}\\
	& p = p_e(r) \left( \frac{\Phi(r,h)-\Phi(r,z)}{\Phi(r,h)-\Phi(r,0)} \right)^{n+1}. \label{pDep}
\end{align}  
Assuming the height is small compared to the radius $|z|\ll r$ we may get the usually used (\cite{Hoshi77,Matsumoto84}) dependence $\rho=\rho_e(r)(1-z^2/h^2)^n$. However, we did not make such an assumption later on. 

We will need integrals from such vertical dependencies:
\begin{align}
	\zeta_n(r,h) = \int_{-h}^h \left( \frac{\Phi(r,h)-\Phi(r,z)}{\Phi(r,h)-\Phi(r,0)} \right)^n dz. \label{intDef}
\end{align}
Apparently there is no general expression for such an integral for arbitrary $n$, but it may be calculated for particular values.

\section{Generalised Rankine-Hugoniot relation} \label{R-H}
Let us denote by index 1 the values in TL and by index 2 the values in the disc. 
Then the mass continuity condition leads to:
\begin{align}
	\rho_{e1}u_1 \zeta_{n,1} = \rho_{e2}u_2 \zeta_{n,2} = j, \label{massCont}
\end{align}
where, for the sake of brevity, we have redefined (\ref{intDef}) as $\zeta_{k,i}=\zeta_{k}(r_2,h_i)$. The momentum continuity:
\begin{align}
	\rho_{e1}u_1^2 \zeta_{n,1} + p_{e1} \zeta_{(n+1),1} = \rho_{e2}u_2^2 \zeta_{n,2} + p_{e2} \zeta_{(n+1),2} \label{momCont}
\end{align}
From (\ref{massCont})-(\ref{momCont}) we may get the following expression for the mass flux:
\begin{align}
	j^2 = \frac{ p_{e2} \zeta_{(n+1),2} - p_{e1} \zeta_{(n+1),1}  }{  [\rho_{e1} \zeta_{n,1}]^{-1} - [\rho_{e2} \zeta_{n,2}]^{-1} }. \label{massContSQ}
\end{align}
The integrated over height energy continuity condition has the following form:
\begin{align}
	h_1 u_1^2 + \zeta_{1,1} w_{e1} = h_2 u_2^2 + \zeta_{1,2} w_{e2}, \label{enCont}
\end{align}
where $w_e$ -- enthalpy at the equator. 
From (\ref{enCont}) taking into account (\ref{massCont})-(\ref{massContSQ}) we may derive the generalization of the Rankine-Hugoniot relation:
\begin{align}
	\frac{h_1 [\rho_{e1}  \zeta_{n,1}]^{-2} - h_2 [\rho_{e2}  \zeta_{n,2}]^{-2} }{[\rho_{e1}  \zeta_{n,1}]^{-1} - [\rho_{e2}  \zeta_{n,2}]^{-1}} = \frac{ \zeta_{1,2} w_{e2} -  \zeta_{1,1} w_{e1}}{p_{e2} \zeta_{(n+1),2} - p_{e1} \zeta_{(n+1),1}}. \label{RHrel}
\end{align}
If we set homogeneous vertical distribution $\zeta_{k,i}=\zeta_k(r_2,h_i)=2h_i$ and $h_1=h_2$ then the usual expression may be gotten: 
\begin{align}
\frac{\rho_{e2}^{-1} + \rho_{e1}^{-1}}{2} = \frac{w_{e2}-w_{e1}}{p_{e2}-p_{e1}}.
\end{align}

For a given equation of state and known values of thermodynamic quantities on both sides of the discontinuity, (\ref{RHrel}) may be considered as an implicit equation relating the heights $h_1$, $h_2$ and the radius $r_2$.
\begin{figure}
	\centering
	\includegraphics[width=1\linewidth]{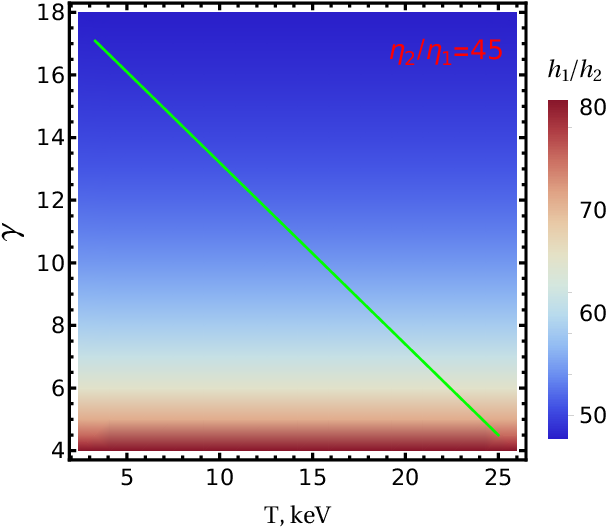}
	\caption{The dependence of the ratio of the TL height to that of  disk  $h_1/h_2$ on the Reynolds number $\gamma_2 = {\dot{M}}/{4\pi h_2 \eta_2}$ and temperature in the TL $T_1$ for the ratio of turbulent viscosities $\eta_2/\eta_1=45$. All other calculations were carried out along the diagonal line, which reflects the transition between the high/soft and low/hard states.  }
	\label{fig:delt45}
\end{figure}

\begin{figure*}[h!]
	\begin{minipage}{0.49\linewidth}
		\centering\includegraphics[width=1\linewidth]{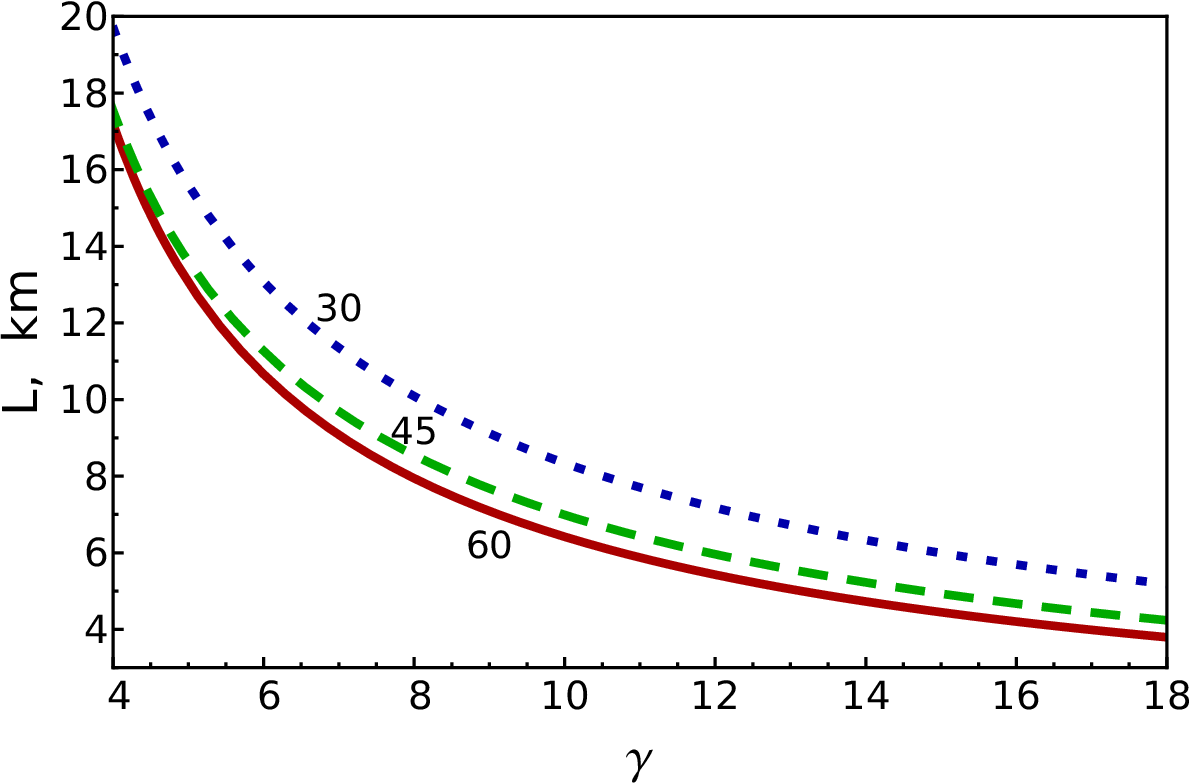}
	\end{minipage}
	\hfill
	\begin{minipage}{0.49\linewidth}
		\centering\includegraphics[width=1\linewidth]{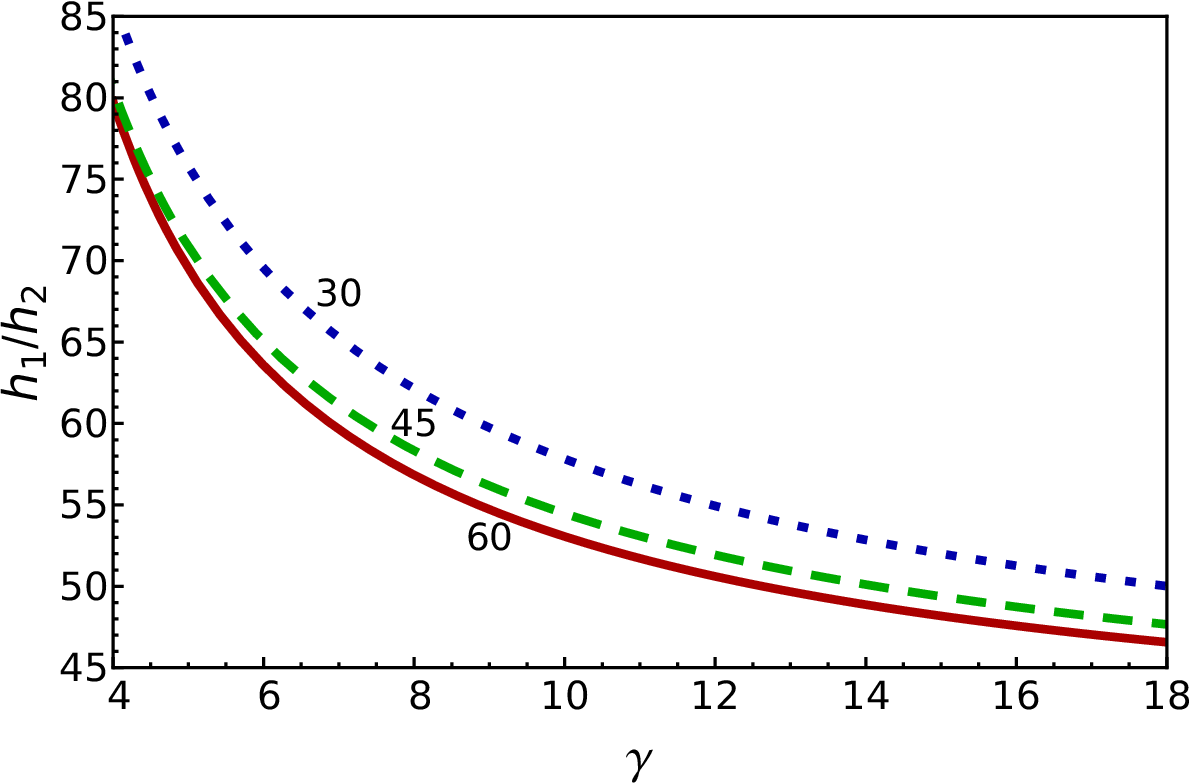}
	\end{minipage}
	\caption{The dependence of the TL length $L$ (left) and the ratio (right) of the TL height to disk height $h_1/h_2$ on the Reynolds number, $\gamma_2 = {\dot{M}}/{4\pi h_2 \eta_2}$ for different ratios of turbulent viscosities $\eta_2/\eta_1$: 60 (solid line), 45 (dashed) and 30 (dotted). }
	\label{fig:lendel}
\end{figure*}

\section{Rotation}
\label{rot}

The continuity equation, integrated over the height has the following solution:
\begin{align}
	r \rho_{e} u \zeta_n = -q,
\end{align}
where $q=\dot{M}/2\pi>0$ is a constant. The equation of force balance in $\phi$ direction for the TL is the following:
\begin{align}
	\frac{\partial }{\partial r} (r^2 \rho_1 u_1 r \omega_1 - \eta_1 r^3 \omega'_1)=0. \label{omEq}
\end{align}
Integrated over the height it has the solution $\omega_1 = C_1 r^{-\gamma_1} + C_2 r^{-2}$, where we denoted 
\begin{align}
	\gamma_1=\frac q {2 h_1 \eta_1} = \frac{\dot{M}}{4\pi h_1 \eta_1},
\end{align}
which is nothing more than the Reynolds number. Before defining the unknown constants, let us discuss the boundary conditions.

In our model, we do not take into account the gradual broadening of the flow after the shock wave, nor its gradual narrowing near the CO. Therefore, neglecting this, we require equality of angular velocities to the disk one at the outer boundary and the CO one at the inner boundary:
\begin{align}
	&\omega_1(r_1)= \omega_s, \label{omR1}\\
	&\omega_1(r_2) = \omega_K(r_2) \label{omR2}.
\end{align}

In addition to the conservation of radial momentum at the discontinuity, there must also be conservation of momentum in the $\phi$-direction. Given the viscosity, the law of $\phi$-momentum continuity, integrated over height, reads as follows:
\begin{align}
	j\, \omega_1 - 2h_1 \eta_1 \omega'_1 = j\, \omega_2 - 2h_2 \eta_2 \omega'_2,
\end{align}
where $j$ is the mass flux from (\ref{massCont}).

Since we suppose that at the outer boundary $r=r_2$ both angular velocities are equal to the Keplerian one (\ref{omR2}), then we have the following condition for the angular velocity derivative: 
\begin{align}
	\omega'_1(r_2) = \frac{\eta_2}{\eta_1} \frac{h_2}{h_1} \omega'_K(r_2). \label{omDerR2}
\end{align} 

Thus, for the second-order differential equation (\ref{omEq}) we have three boundary conditions (\ref{omR1})(\ref{omR2})(\ref{omDerR2}), i.e. the problem is overdetermined. Using the conditions (\ref{omR1})(\ref{omR2}) we may write down the solution:
\begin{align}
	\omega_1 = \frac{1}{r_1^2 r_2^{\gamma_1} - r_1^{\gamma_1} r_2^2}\left(  \frac{r_1^2 r_2^2}{r^2}(\omega_\text{K} r_2^{\gamma_1} - \omega_s r_1^{\gamma_1})  - \frac{r_1^{\gamma_1} r_2^{\gamma_1}}{r^{\gamma_1}}(\omega_\text{K} r_2^2 - \omega_s r_1^2)  \right), \label{omegaSol}
\end{align} 
and (\ref{omDerR2}) may be used to get another implicit equation for $h_1$, $h_2$ and $r_2$ with given $\omega_s$, $\omega_K(r_2)$ and $r_1$. Thus, knowing the CO radius $r_1$ and height of the accretion disk $h_2$, as well as the thermodynamic characteristics of the plasma $\rho_{e}$, $p_{e}$, $w_{e}$ on both sides of the discontinuity, the length, and height of the transition layer can be calculated using (\ref{RHrel})(\ref{omDerR2})(\ref{omegaSol}).

\section{Setup and results}\label{setup}

Since we aim to calculate the geometric characteristics $h_1$, $r_2$ of  the TL, in (\ref{omegaSol}) we have switched to the Reynolds number for the disk $\gamma_2$, which is independent of $h_1$:
\begin{align}
	\gamma_2 = \gamma_1 \left( \frac{\eta_1}{\eta_2} \right) \left(\frac{h_1}{h_2}\right) = \frac{\dot{M}}{4\pi h_2 \eta_2}.
\end{align}

As the CO we considered an NS with a mass of $M=1.5 M_\odot$, a radius of $r_1=15\text{ km}$ and an angular velocity $\omega_s = 10^{-2}\text{ s}^{-1}$. The gravitation potential of the NS is the Newtonian,  $\Phi=-GM/(r^2+z^2)^{1/2}$.

Although in deriving the vertical distributions (\ref{rhoDep})-(\ref{pDep}) it was assumed that the gas is polytropic, the equation of state for the quantities at the equator can, generally speaking, be given arbitrarily; but we considered the same equation of state for vertical and radial dependences. For both TL and accretion disk we took the polytropic index $n=3$, which corresponds to the photon gas equation of state with $w=4p/\rho$ and $p=\alpha T^4/3$, where $\alpha$ -- radiation density constant, $T$ -- temperature. 

As a disk height, we took three Schwarzchild radii $h_2=3r_{\rm S} =13.3\text{ km}$. The maximum (at the equator) density and temperature in the disk were taken as $\rho_{e2}=1.2 \cdot 10^{-5}\text{ g cm}^{-3}$ and $T_{e2}=0.5\text{ keV}$ {see e.g. ST09}.

The density for the TL was chosen to correspond to the observed optical depth relative to Thompson scattering $\tau\simeq 2$ (see e.g. ST09). This value can be achieved by selecting the concentration $10^{17}\text{ cm}^{-3}$ which gives $\rho_{e2}=2 \cdot 10^{-7}\text{ g cm}^{-3}$.

For neutron stars, the TL temperature is around $25$ keV in the low/hard state and evolves to approximately $2$ keV in the high/soft state (e.g. \cite{Seifina10}). In our analysis, there is no way that we can establish a law of temperature change as a function of the mass accretion rate. Therefore, to begin with, we solved (\ref{RHrel})(\ref{omDerR2})(\ref{omegaSol}) for the entire temperature range. As can be seen from Fig.~\ref{fig:delt45} the height of the TL is almost independent of the temperature in it. The same picture is observed for the TL length. However, the solution of  (\ref{RHrel}) with some fixed $r_2$ alone shows considerable sensitivity to the temperature. The independence of the geometry of the TL on its temperature obtained in our model is achieved only by solving the equations for height and length together. Thus, we can conclude that the geometric characteristics of  the TL depend mainly on the accretion rate (Reynolds number $\gamma$). Hereafter, these geometric characteristics have been calculated assuming a simple linear relation, $T_1(\gamma_2)$, despite the insignificance of this correction. The relation, $T_1(\gamma_2)$, represented in Fig.~\ref{fig:delt45}, reflects the described above transition between the high/soft and low/hard states. 

The turbulent viscosity may be estimated in different ways. SS73 proposed its value as $\eta = \alpha \rho c_s l_\text{turb}$, where  $\alpha$ is a dimensionless constant, $c_s$ is the sound speed, and $l_\text{turb}$ is a turbulent length scale, which is equal to the height $h$ or, in addition, to the value $(h^{-2} +h_r^{-2})^{-1/2}$, where $h_r$ is the radial pressure scale $h_r=|p/p'|$ (\cite{Popham01}, PS01 hereafter). Also in radiation-pressure dominated
conditions it may be estimated (TLM98) as $\eta=m_p n_\text{ph}c l/3$, where $m_p$  is the proton mass, $n_\text{ph}$ is the photon number density,  $c$ is the speed of light, and $l$ is the photon mean free path. So the turbulent viscosity may depend on the conditions in the flow as well as its geometry. We avoided further reduction and took several values of the ratio of turbulent viscosities $\eta_2/\eta_1$ to make the resulting TL lengths $L$ to be near the observable ones as well as  $\gamma_2$  corresponds to $\alpha^{-1}$ parameter from the SS73 model.

We calculated (\ref{RHrel})(\ref{omDerR2})(\ref{omegaSol}) with parameters describes above for three different values of the ratio $\eta_2/\eta_1$: 30, 45, and 60. When this value is increased, there is no significant change, whereas, at values quite below 30, the length of TL becomes absurdly large. 

As can be seen from Fig.~\ref{fig:lendel}, the TL length decreases with increasing $\gamma_2$. At extremely low accretion rates, the length of the TL considerably exceeds the radius of the CO, and as it increases, it can drop to tenths of it. The motion of a gas element is a superposition of rotation with an angular velocity $\omega$ and falling on the CO with the velocity $u$, i.e.\@ it is a spiral motion. A smaller radial gas flow (i.e.\@ smaller $\dot{M}$ and $\gamma_2$) increases the distance between the coils of such a spiral, thereby making it larger. Therefore, with a fixed viscosity responsible for decreasing the angular momentum, increasing the mass accretion rate leads to decreasing the TL length.

The behavior of the TL height $h_1$ is similar -- it is about 80 times the height of the disk at a low accretion rate and drops to about 45 when the accretion rate increases (see Fig.~\ref{fig:lendel}). Thus, the height of the TL of an NS can be up to a thousand kilometers in the low/hard state and half that in the high/soft state. In the frame of our model, this result is quite explainable.  The fact is that an increase in the mass accretion rate at a fixed density $\rho_2$ implies an increase in radial velocity. In this way, more matter flows through a certain area per unit of time, making it smaller.

\section{Conclusion}\label{conclusion}
We have solved  the problem of  the transition layer  geometry  near a central object (CO).  It was supposed to be weakly magnetized, therefore we study only the hydrodynamical  properties of such configurations. We have derived the generalized Rankine-Hugoniot relation taking into account different the heights of the accretion disk and TL and heterogeneous distributions along them. Then we solved the equations of continuity and force balance in $\phi$ direction and have shown that the last one ought to have three boundary conditions. In order to satisfy them, a TL must have certain geometric characteristics.  

As a CO we considered NS, setting the gas parameters on both sides of the shock wave according to observations (see ST09). Then the length and height of the transition layer were calculated. We suggest that the turbulent  viscosity is constant on each side of the shock,  which values have been  chosen as such that the Reynolds number $\gamma_2$  corresponds to $\alpha^{-1}$ parameter from the SS73 model. As can be seen from Fig.~\ref{fig:lendel}, both the TL length and height decrease with increasing $\gamma_2$. Therefore we may conclude that:

1. We clarify  the nature of the region where X-ray spectra are formed.

2.  When the source evolves to a softer state, the Compton corona region becomes more compact (see Fig. \ref{fig:lendel} and also  ST06, ST09).  We  should emphasize that  the integrated power $P_x$  of the resulting power density spectra rapidly declines toward soft states (TSA07).

3. We should point out  that a ratio of the TL height  to the disk height $h_1/h_2$ on the Reynolds number is quite  large  (see Fig. \ref{fig:lendel}).

4. The calculations were done for  a NS, although the same behavior should be expected for other types of COs (a WD and a BH).

5. According to the presented calculations and observed manifestations (ST06, ST09) the TL height $h_1$ sufficiently exceeds the disk height $h_2$. Therefore, as also noted in PS01, one-dimensional constant-height models need to be improved to correctly describe the transition layer.


\bibliographystyle{aa}

\centering\textsc{Make science, not war } \bcpeaceandlove

\end{document}